\documentclass{PoS}

\title{Parameter Tuning of Three-Flavor Dynamical Anisotropic Clover Action}

\ShortTitle{Tuning 3f Anisotropic Clover Action}

\author{\speaker{Huey-Wen Lin}\\
        Thomas Jefferson National
Accelerator Facility, Newport News, VA 23606\\
        E-mail: \email{hwlin@jlab.org}}

\author{Robert G. Edwards  and B\'alint Jo\'o\\
        Thomas Jefferson National
Accelerator Facility, Newport News, VA 23606\\}

\abstract{In this work, we perform parameter tuning with dynamical anisotropic clover lattices using the Schr\"odinger functional and stout-smearing in the fermion field. We find that $\xi_R/\xi_0$ is relatively close to 1 in our parameter search, which allows us to fix $\xi_0$ in our runs. We proposed to determine the gauge and fermion anisotropy in a Schr\"odinger-background small box using Wilson loop ratios and PCAC masses. We demonstrate that these ideas are equivalent to but more efficient than the conventional meson dispersion approach. The spatial and temporal clover coefficients are fixed to the tree-level tadpole-improved clover values, and we demonstrate that they satisfy the nonperturbative condition determined by Schr\"odinger functional method.
}

\FullConference{The XXV International Symposium on Lattice Field Theory\\
		 July 30-4 August 2007\\
		 Regensburg, Germany}

\begin{document}

\vspace{-1cm}
\section{Introduction}
\vspace{-0.5cm}
Lattice quantum chromodynamics (QCD) has successfully calculated many of meson spectrum; however, there remain many challenges for the lattice community to resolve the myriad states present in QCD. The case of the nucleon spectrum is one such battlefield. Consider the lowest three states in the $N$ spectrum ($N$, $N^\prime$ ($P_{11}$) and $N^*$ ($S_{11})$), for example. Many earlier quenched lattice QCD calculations\cite{all-roper}
find a spectrum inverted with respect to experiment, with $N^\prime$ heavier than the opposite-parity state $N^*$. Although the Kentucky group~\cite{Mathur:2003zf} managed to find the correct mass ordering around pion mass 300--400~MeV (after taking care of the effects of the quenched ``ghosts''), no other lattice group has been able to reproduce the experimental ordering using different approaches. Furthermore, these are just the lowest few states in the $N$ spectrum. There are many more states seen in experiment for which lattice calculations could help to identify particle properties.

This situation suggests an urgent need for full-QCD simulations that can resolve some of these issues. In order to get better signal for the excited states (especially for the higher-excited nucleon spectrum), one needs a lattice with a fine temporal lattice spacing. At the same time, we also want to avoid finite-volume effects. Current dynamical lattice gauge ensembles manage to have a reasonable lattice box with spatial dimensions about 3~fm, but unfortunately the lattice spacing is about 2~GeV, which is not fine enough to allow determination of more than one excited state. One solution to this situation would be to generate anisotropic dynamical lattices.

The anisotropic lattice has been widely adopted in lattice calculations. It was first used to simulate heavy-quark physics, such as charm, back in the era when the lattice spacings were too coarse to use the relativistic quark action to simulate heavy quarks. Another main application is for calculations, such as glueballs\cite{Morningstar:1999rf} and multiple excited-state extraction\cite{Basak:2006ww}, where the anisotropic lattice has great improvements over isotropic due to the finer lattice time spacing. Previous results on the anisotropic lattice include two-flavor anisotropic dynamical simulations done by CP-PACS\cite{Umeda:2003pj} and TrinLat collaboration\cite{Morrin:2006tf}.

In this work, we will use a three-flavor Sheikholeslami-Wohlert (clover) action with stout-link smearing (in the spatial direction only) in the Schr\"odinger functional scheme\cite{alpha}. We can determine the gauge anisotropy by looking at the Wilson loop ratios. The ratio of the PCAC-current quark mass with background field in space and in time tells us about the fermion anisotropy, which has been determined in the past using the meson dispersion relation in large volume. The rest of the coefficients are set to tree-level tadpole improved values, with the tadpole factors determined from numerical simulation. The clover coefficients are fixed at tree-level tadpole-improved values, which later we demonstrate are consistent with nonperturbative ones determined in the Schr\"odinger functional scheme. Our configurations have been generated using the Chroma\cite{Edwards:2004sx} HMC
code with RHMC for the third flavor and multi-timescale integration. For more details, see Ref.~\cite{Edwards:2007}.

\vspace{-0.4cm}
\section{Methodology and Setup}
\vspace{-0.3cm}
\subsection{Action}
\vspace{-0.3cm}
We use tree-level tadpole-improved $O(a^2)$-improved Symanzik gauge action:
\vspace{-0.3cm}
\begin{eqnarray}\label{eq:aniso_syzG}
S_G^{\xi} &=& \frac{\beta}{N_c} \left\{ \frac{u_t}{\xi_0 u_s^3}
\sum_{x, s>s^\prime} \left[ c_0{\cal P}_{ss^\prime}+c_1{\cal R}_{ss^\prime}\right]
+
\frac{\xi_0}{u_s^4} \sum_{x,s}\left[ c_0{\cal P}_{st}+c_1{\cal R}_{st}\right] \vphantom{\frac{1}{\xi}} \right\},
\end{eqnarray}
\vspace{-0.2cm}
where the $\xi_0$ is the gauge anisotropy. We adopt the clover fermion action
\vspace{-0.cm}
\begin{eqnarray}
a_t Q_F & = & \frac{1}{u_t} \left\{
 u_t \hat{m_0} + \nu_t \hat{W}_t +\frac{\nu_s}{\xi_0}  \sum_s  \hat{W}_s -\frac{1}{2} \left[
      c_{\rm SW}^t \sum_{s} \sigma_{ts} \hat{F}_{ts} +
      \frac{c_{\rm SW}^s}{\xi_0}
      \sum_{s<s\prime} \sigma_{ss\prime} \hat{F}_{ss\prime} \right]
      \right\}. 
\end{eqnarray}
The $\nu_t$ is a redundant parameter and set to 1 to resolve the doubler problem of Wilson-type fermions. The $\nu_s$ reflects the ratio between fermion and gauge anisotropy. The $c_{\rm SW}^{s,t}$ are the spatial and temporal clover coefficients which are set as follows:
\vspace{-0.2cm}
\begin{eqnarray}
c_{\rm SW}^s = \frac{\nu_s}{u_s^3}, \quad c_{\rm SW}^t =
\frac{1}{2}\left(\nu_t + 
\frac{1}{\xi}\right)\frac{1}{u_t u_s^2};
\end{eqnarray}
these selections are discussed in Ref.~\cite{Chen:2000ej}. The tadpole factors are later set to fixed values taken from our early dynamical simulations, which agree amongst themselves within 1-2\%. Thus we are left to tune the remaining four coefficients: $\xi_0$, $\nu_s$, $m_0$, $\beta$.

\vspace{-0.2cm}
\subsection{Stout-smeared links}
\vspace{-0.2cm}
We will use three-dimensionally stout-smeared link variables\cite{Morningstar:2003gk} within the fermion action. Note that the smearing does not involve the time direction, so the transfer matrix remains physical. As with other smearings, one should check the smearing parameter carefully to avoid potentially incorrect short-distance physics. In this work, we set $\rho=0.22$ and $n_\rho=2$ for exploratory study.

With a nonperturbative determination of the clover coefficients at the target lattice spacing of $a=0.1\mbox{ fm}$, the scaling violations are about 1\% in $a m_V / \sqrt{a^2\sigma}$~\cite{Edwards:1997nh}. These previous scaling studies used an isotropic quenched action. On the left-hand side of Figure~\ref{fig:scaling} (where all the points on the graph with fixed $m_\pi/m_\rho=0.7$), we show the result from CP-PACS's isotropic Wilson scaling behavior; our simulation measurement on the anisotropic lattice shows a more continuum-like scaling. When we use stout-link smearing in the Wilson fermion action, we notice dramatic improvement. In the clover case (on the right), similar tests are performed, compared with nonperturbative clover coefficients.  The stout-link smeared clover action point is compatible with the scaling.

\vspace{-0.2cm}
\subsection{Schr\"odinger functional}
\vspace{-0.2cm}
The Schr\"odinger functional\cite{alpha} has been implemented in lattice QCD since the 1990's. It allows us to simulate at lighter pion mass (since the background field lifts zero modes); from the PCAC relation we can check how close our $c_{\rm SW}$ in the fermion action is compared with the nonperturbative value.

We modify our definition of the quark mass to be
\vspace{-0.2cm}
\begin{eqnarray} \label{eq:M_redef}
M(x_0,y_0) &=& r(x_0) - \frac{{r^\prime(y_0)-r(y_0)}}{{s^\prime(y_0)-s(y_0)}} s(x_0)
; \,\,\, M^\prime(x_0,y_0) 
=r^\prime(x_0)-\frac{{r^\prime(y_0)-r(y_0)}}{{s^\prime(y_0)-s(y_0)}} s^\prime(x_0),
\end{eqnarray}
where $r$ and $s$ are obtained from
\vspace{-0.2cm}
\begin{eqnarray}
r(x_0) & = &  0.25  \left( \partial_0 + \partial_0^* \right)
f_A(x_0) / f_P(x_0) 
; \,\, \,
s(x_0)  =0.5 a \, \partial_0 \partial_0^* f_P(x_0) /f_P(x_0).
\end{eqnarray}
$f_{A(P)}$ (with $\Gamma=\gamma_5\gamma_\mu (\gamma_5)$ ) is a correlation function of bulk fields ($\overline{\psi}$, $\psi$) and boundary fields at $t=0$ ($\overline{\eta}$, $\eta$):
\vspace{-0.2cm}
\begin{eqnarray}
f_{O_\Gamma} (t)&=& \langle \overline{\psi}\Gamma\psi({\bf x},t) \sum_{\bf y,z} \overline{\eta}({\bf y},t)\Gamma\eta({\bf z},t) \rangle/(N_f^2-1).
\end{eqnarray}
Similar definitions apply to $r^\prime$ and $s^\prime$, but the $f_{A(P)}^\prime$ now involves the other boundary fields at $t=T,T-1$ ($\overline{\eta}^\prime$, $\eta^\prime$) and a sign change. Later in this work, we will calculate the modified quark mass to tune our coefficients.

On the isotropic lattice, one can also determine the $c_{\rm SW}$ coefficient in Schr\"odinger functional scheme from the PCAC relation. We require the nonperturbative value of $c_{\rm SW}$ to lie at
\vspace{-0.2cm}
\begin{eqnarray}\label{eq:np_csw}
\Delta M = M(x_0,y_0)-M^\prime(x_0,y_0) = \Delta M^{(0)},
\end{eqnarray}
where $\Delta M^{(0)}$ is the tree-level mass splitting obtained from a classic background field simulation with the same setup of gauge and fermion actions. Previous dynamical works were carried out by Alpha with $N_f=2$\cite{Jansen:1998mx} using Wilson gauge only, CP-PACS for two-flavor and three-flavor calculations with Wilson and Iwasaki gauge actions\cite{Umeda:2003pj,Yamada:2004ja,Aoki:2005et}. However, all applications of the Schr\"odinger functional so far have been limited to isotropic lattices. This work is the first to apply the Schr\"odinger functional to dynamical anisotropic lattices.

\begin{figure}
\begin{tabular}{cc}
\begin{minipage}{0.55\textwidth}
\vspace{-0.7cm}
\includegraphics[width=1.\textwidth]{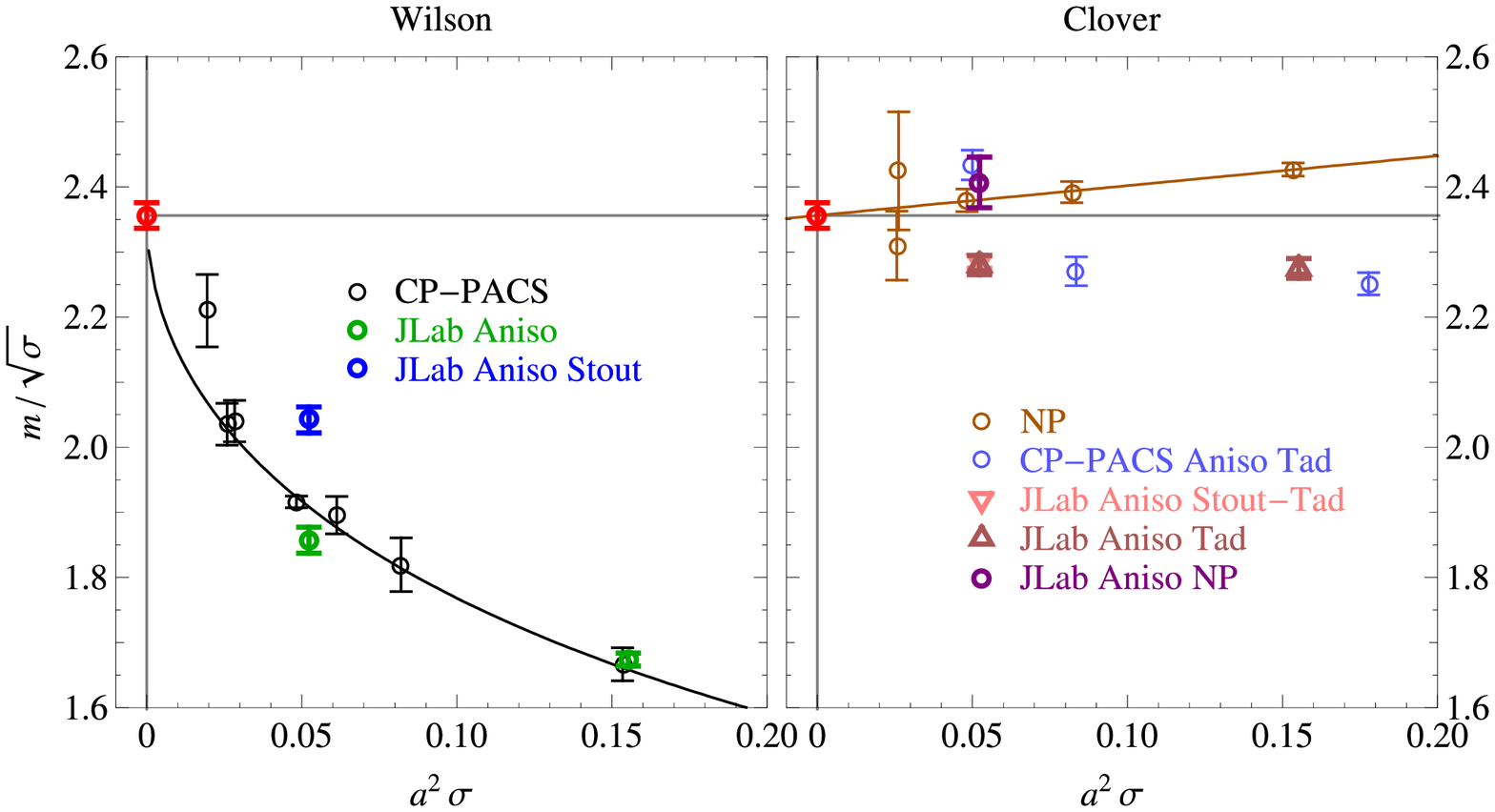}
\caption{
The scaling behavior of quenched Wilson gauge action with Wilson (left) and clover (right) fermion actions.}
\label{fig:scaling}
\end{minipage}
&
\begin{minipage}{0.45\textwidth}
\vspace{-1cm}
\includegraphics[width=0.48\textwidth]{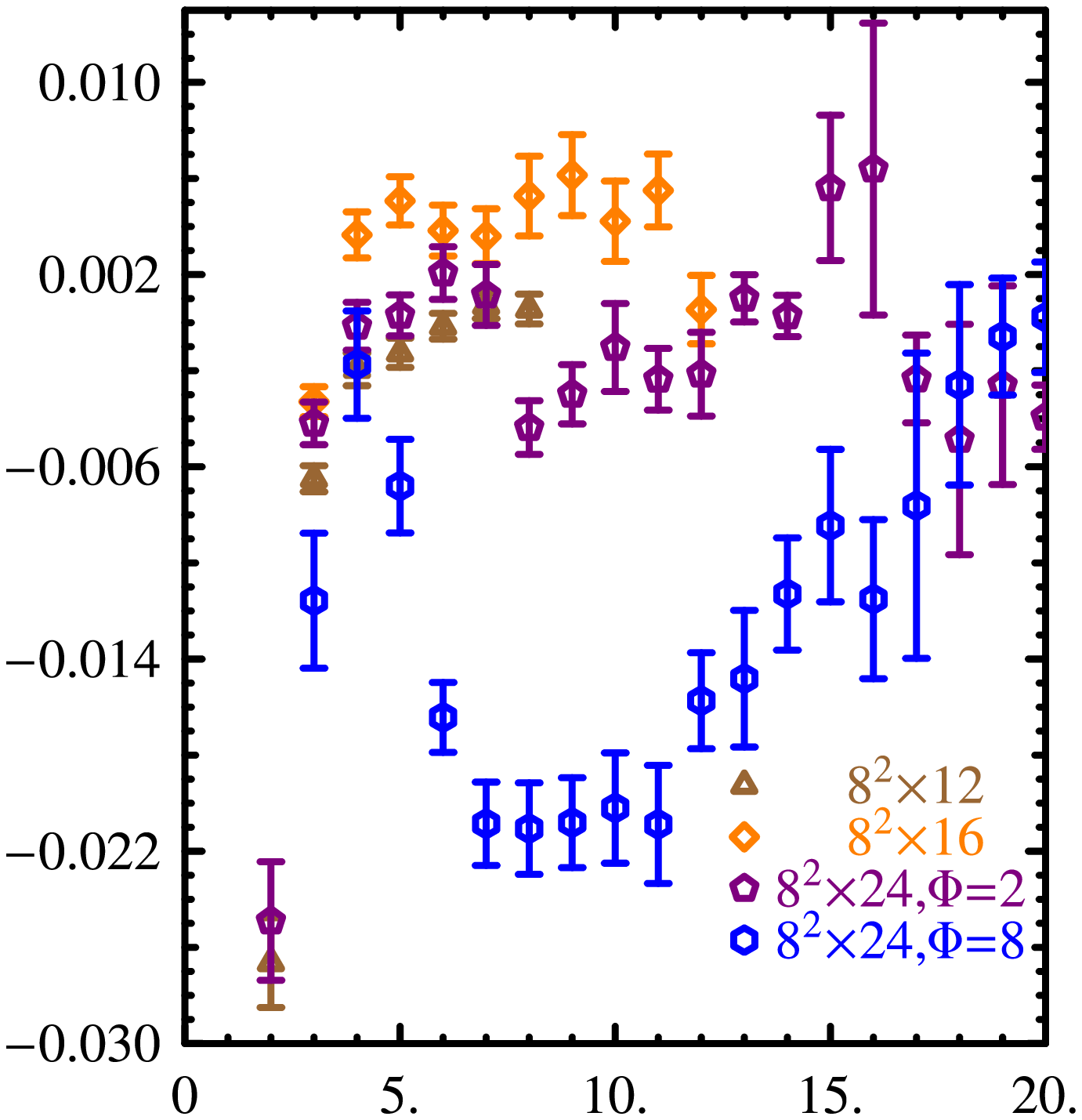}
\includegraphics[width=0.48\textwidth]{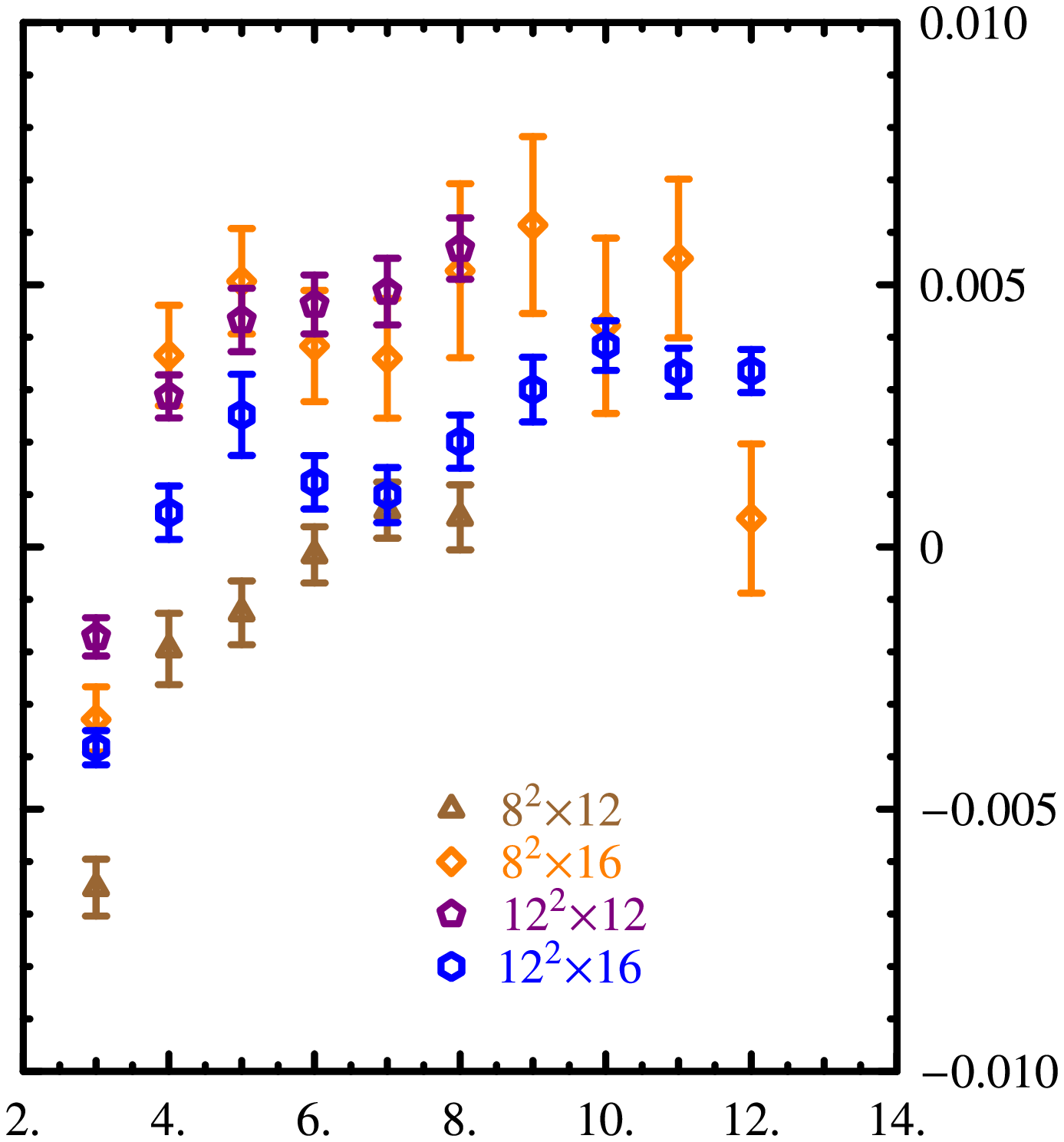}
\caption{$a_s M_s$ comparison among different spatial box volumes
}\label{fig:size-test}
\end{minipage}
\end{tabular}
\end{figure}

\begin{figure}
\vspace{-0.2cm}
\includegraphics[width=0.5\textwidth]{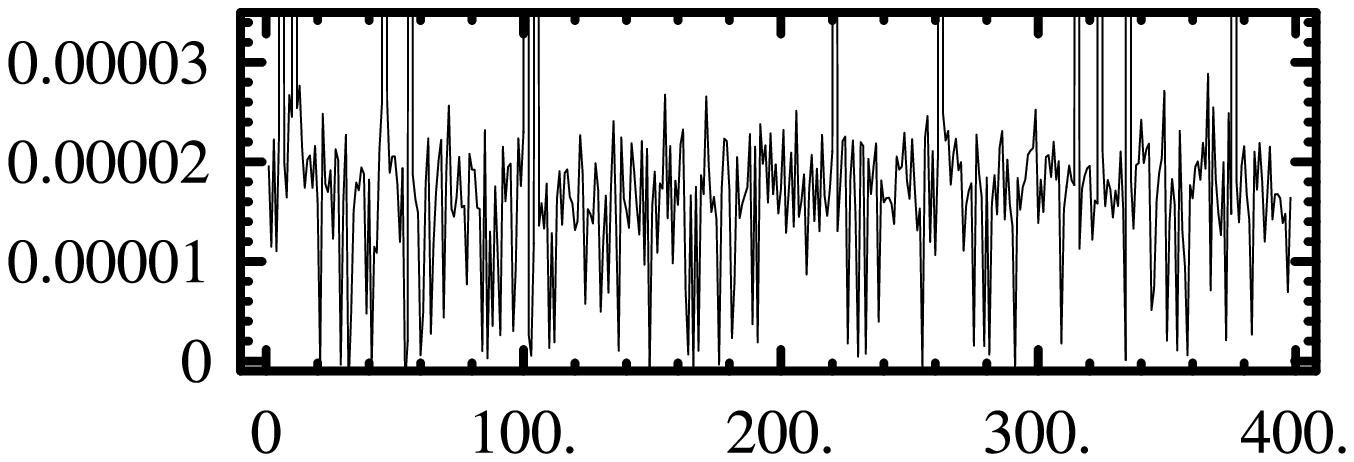}
\includegraphics[width=0.5\textwidth]{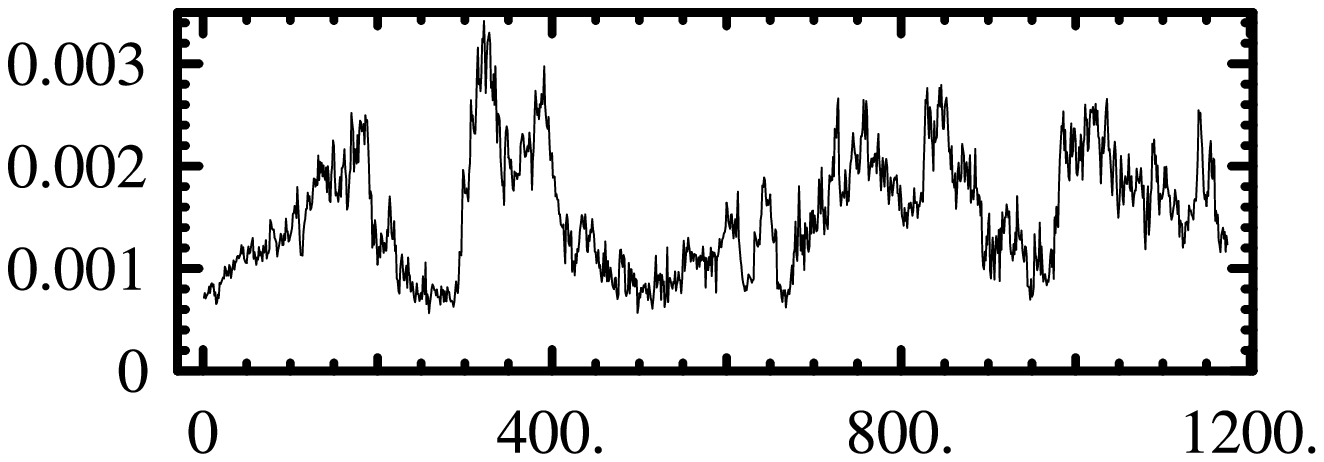}
\vspace{-0.2cm}
\caption{$\lambda_{\rm min}(Q^2)$ measured in simulation with (left) and without (right) a background field. The $x$-axis is in units of 5 trajectories.}\label{fig:lambda}
\end{figure}

In this work, we will implement the Schr\"odinger functional setup on {\it anisotropic} lattices for the first time in dynamical simulations. We find that in the three-flavor (anisotropic) clover action (with parameter $m_0=-0.054673$, $\nu_s=1$) simulation, the lowest eigenvalue of $Q^\dagger Q$ is lifted by the background field. This helps to reduce the frequency of exceptional eigenvalues, as shown in Figure~\ref{fig:lambda}. The eigenvalue changes dramatically from 0--6000 trajectories without the background field (right figure). A reduction in near-zero eigenvalues gives us better acceptance rate in the HMC.

We also implement the background field not only in the ``$t$'' direction (as conventionally used in Schr\"odinger functional) but also the ``$z$'' direction. Less is known about putting the background field in the ``$z$'' direction. (We label the modified masses $M_s$ and $M_t$ with respect to the direction of the boundary field direction and similarly for the mass difference $\Delta M_s$ and $\Delta M_t$.) We measure $M(x_0)$ with various $L_z$, showing results in Figure~\ref{fig:size-test}. When we increase the length in the $z$ direction, good signal appears from 12 to 16 but not beyond 24. This is because the background field becomes too weak at large $L_z$. When we increase the background field signal, which is proportional to $\Phi$ at $L_z=24$, the signal is still not as good as for $L_z=16$. Similar checks can be done regarding the size of $L_{x,y}$. As we increase the value from 8 to 12, the signal shows improvement. The right panel of Figure~\ref{fig:size-test} shows an optimal choice of $12^3$ for the spatial volume; this is what we use in the remainder of this work.

\vspace{-0.2cm}
\subsection{Proposed conditions}
\vspace{-0.2cm}
In summary, we implement Schr\"odinger functional with background fields in two directions: ``t'' and ``z'', and we measure the quantities $M_{t,s}$ and $\Delta M_{s,t}$. We will determine the gauge anisotropy $\xi_R$ from the ratios of static quark potential (more details in the following section). We will get the fermion anisotropy $\nu_s$ by measuring the PCAC mass ratio in finite-volume Schr\"odinger functional scheme; in our coefficient definition, we are looking for $a_sM_s/a_tM_t=\xi_R$. We set the two clover coefficients ($c_{\rm SW}^{s,t}$) to the stout-smeared tadpole  coefficient, where the tadpole factor is numerically tuned. We further check the nonperturbative conditions in Eq.~\ref{eq:np_csw} from our measured the PCAC mass difference $\Delta M_{s,t}$; we find that they are consistent.

\vspace{-0.2cm}
\section{Numerical Results}
\vspace{-0.4cm}
In this work, we fix $\beta=2$ for an exploratory study on the tuning of action parameters.

\begin{figure}
\begin{tabular}{cc}
\begin{minipage}{0.47\textwidth}
\vspace{-0.8cm}
\includegraphics[width=1\textwidth]{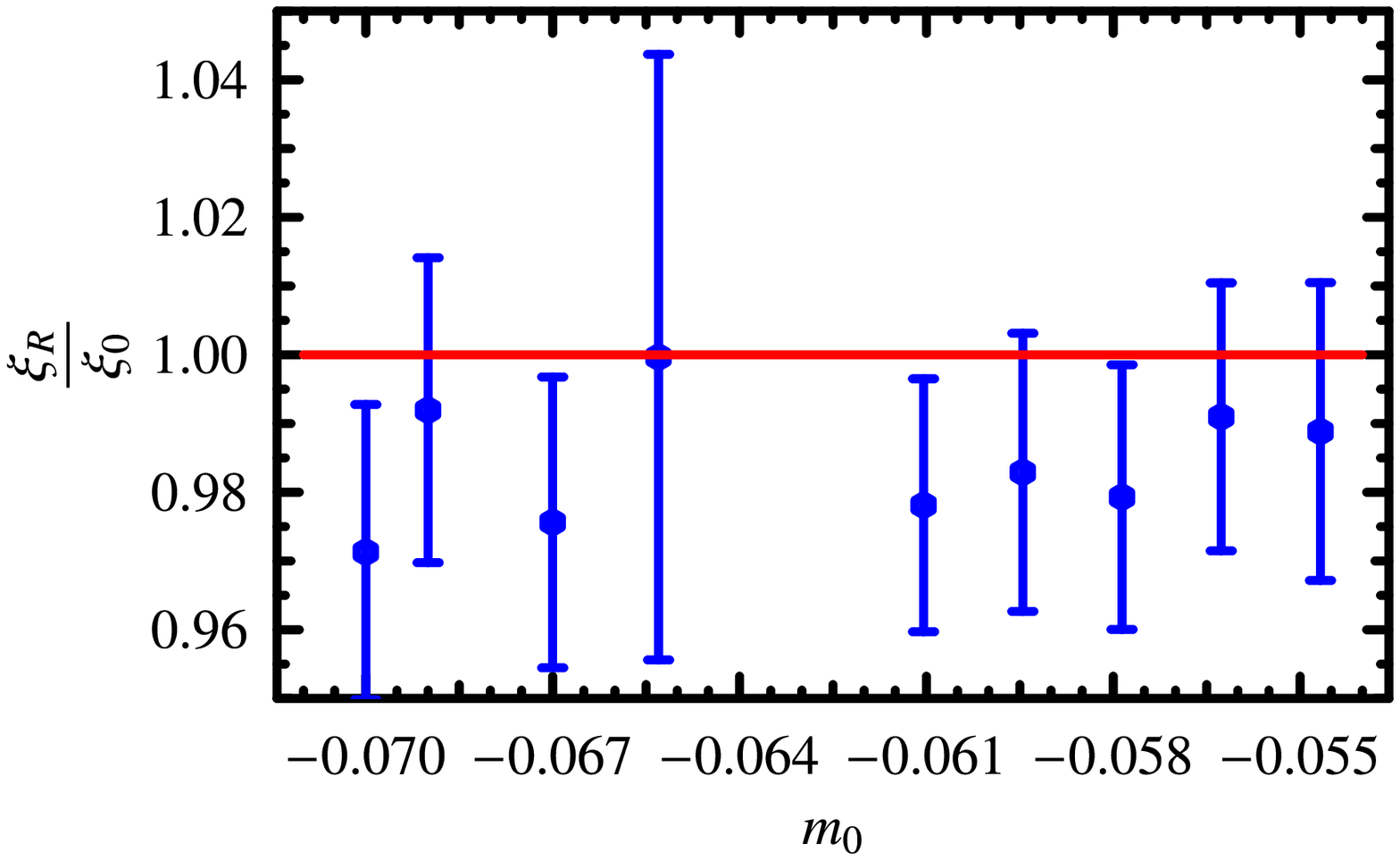}
\vspace{-0.99cm}
\caption{$m_0$ dependence of the ratio $\xi_R/\xi_0$ for $\nu=1$ }\label{fig:xiRxi0}
\end{minipage}
&
\begin{minipage}{0.49\textwidth}
\vspace{-0.8cm}
\includegraphics[width=\textwidth]{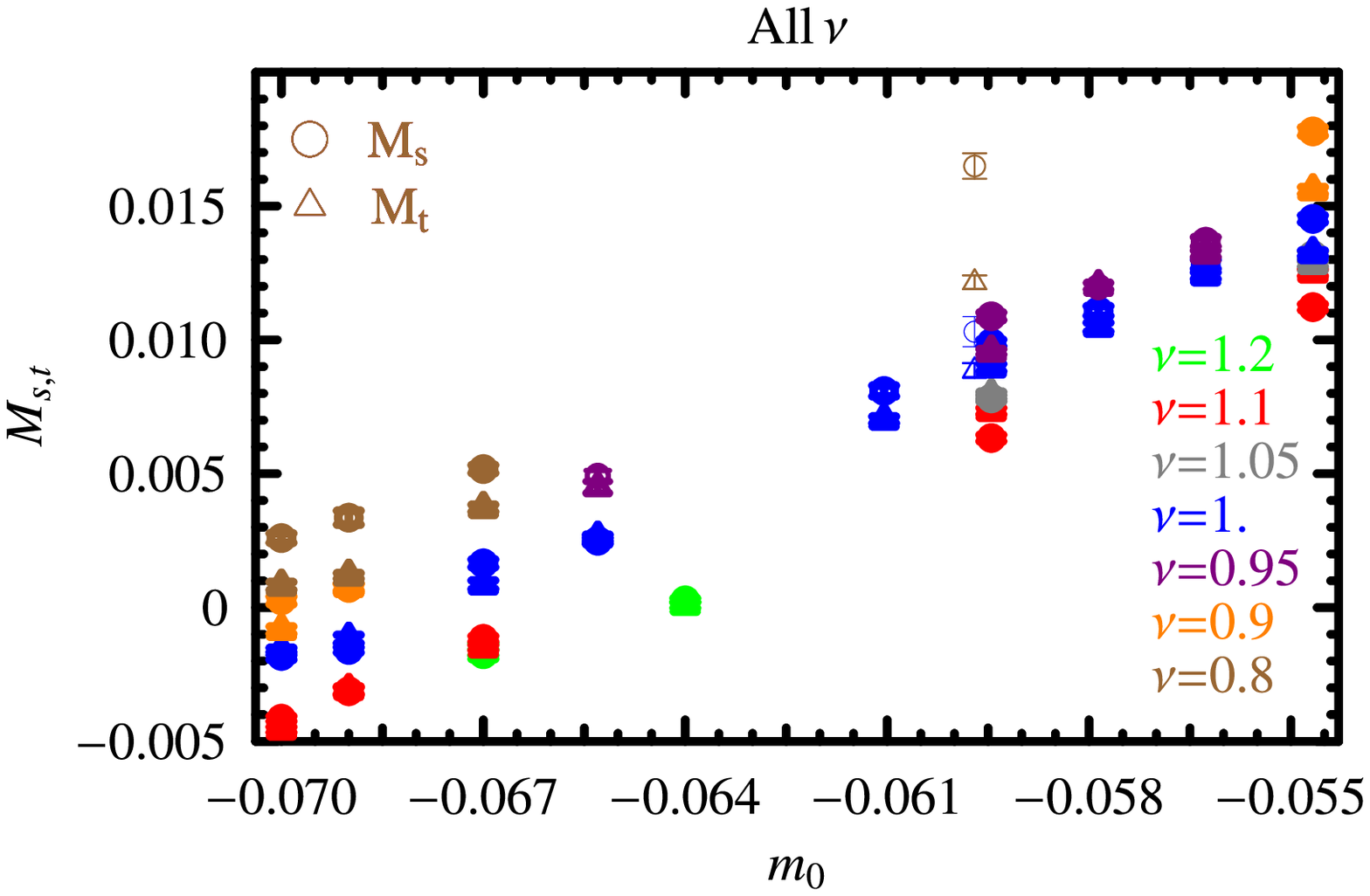}
\caption{The measured $M_{s,t}$ (in units of $a_t^{-1}$)
at each $\nu$ value as a function of $m_0$
}\label{fig:mst}
\end{minipage}
\end{tabular}
\end{figure}

The natural way to look for anisotropy in the gauge sector is to start with the static quark potential. We calculate ratios of Wilson loops involving the temporal direction $W_{st}$ and those without it $W_{ss}$\cite{Klassen:1998ua,Umeda:2003pj}:
\begin{eqnarray}\label{eq:WL_ratio}
R_{ss}(x,y) &=&  \frac{W_{ss}(x,y)}{ W_{ss}(x+1,y)}
\rightarrow_{\rm asym.} e^{a_s V_s (y a_s)} ; \,\,\,
R_{st}(x,t) =  \frac{W_{st}(x,t)}{W_{st}(x+1,t)}
\rightarrow_{\rm asym.} e^{a_s V_s (t a_t)}.
\end{eqnarray}
The finite-volume differences in the ratio $R_{st}$ and $R_{ss}$ are the same if $N_t=\xi_R N_s$. Thus, there are no finite-volume differences in $V_s(y a_s)$ or $V_s(t a_s/\xi_R)$ either. Naturally, one should impose $R_{ss}(x,y)\stackrel{!}{=}
R_{st}(x,t)$ to get the renormalized $\xi_R$. The $\xi_R/\xi_0$ is consistent with 1 within a few percent from our parameter searching. (Figure~\ref{fig:xiRxi0} shows a special case when $\nu_s=1$). We can set $\xi_0=3.5$ and tune only the remaining two parameters.

\begin{figure}
\begin{tabular}{cc}
\begin{minipage}{0.49\textwidth}
\vspace{-0.8cm}
\includegraphics[width=\textwidth]{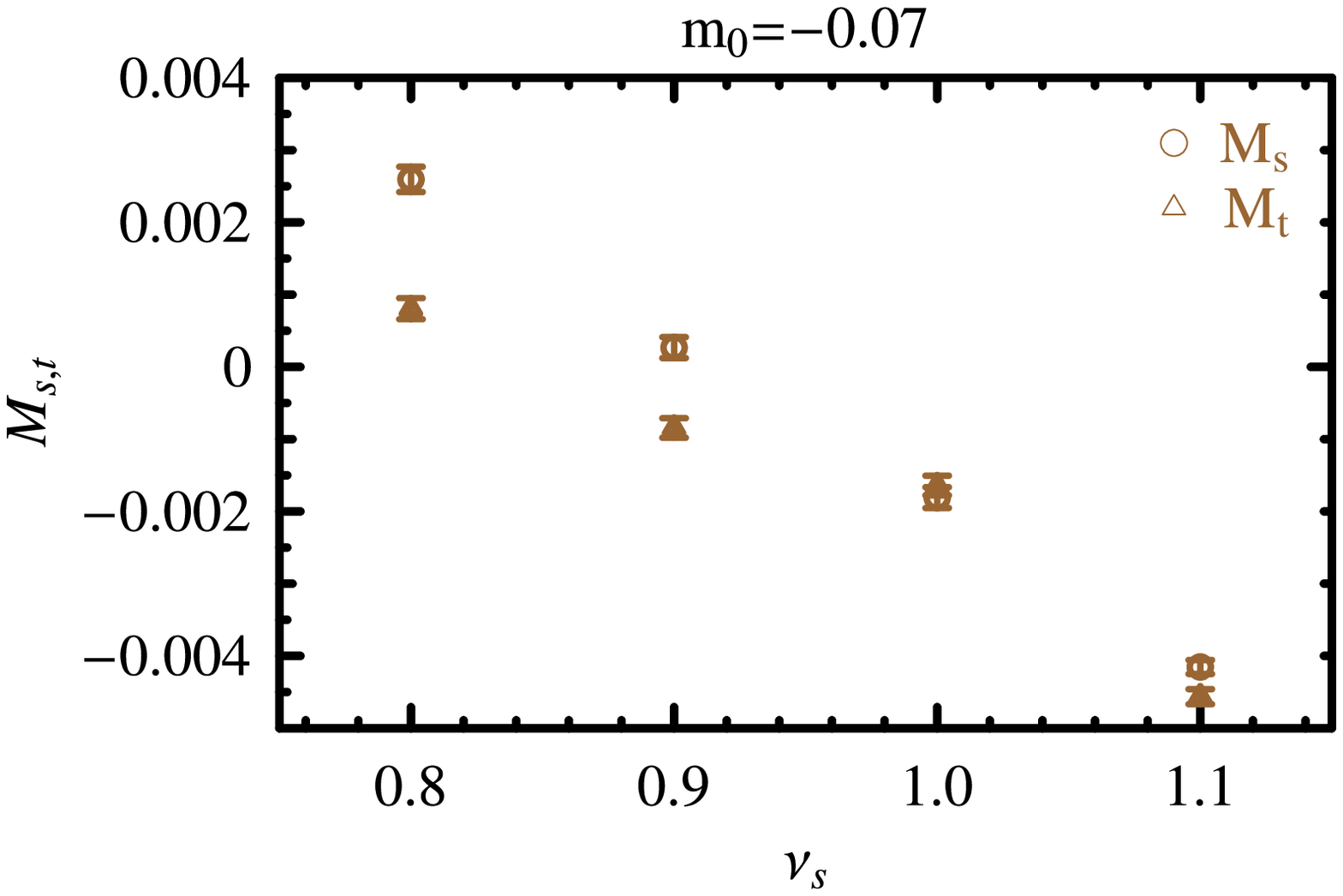}
\caption{The measured $M_{s,t}$ in units of $a_t$  as a function of $\nu$ at fixed $m_0=-0.07$
}\label{fig:m0-07}
\end{minipage}
&
\begin{minipage}{0.47\textwidth}
\vspace{-0.9cm}
\includegraphics[width=0.99\textwidth]{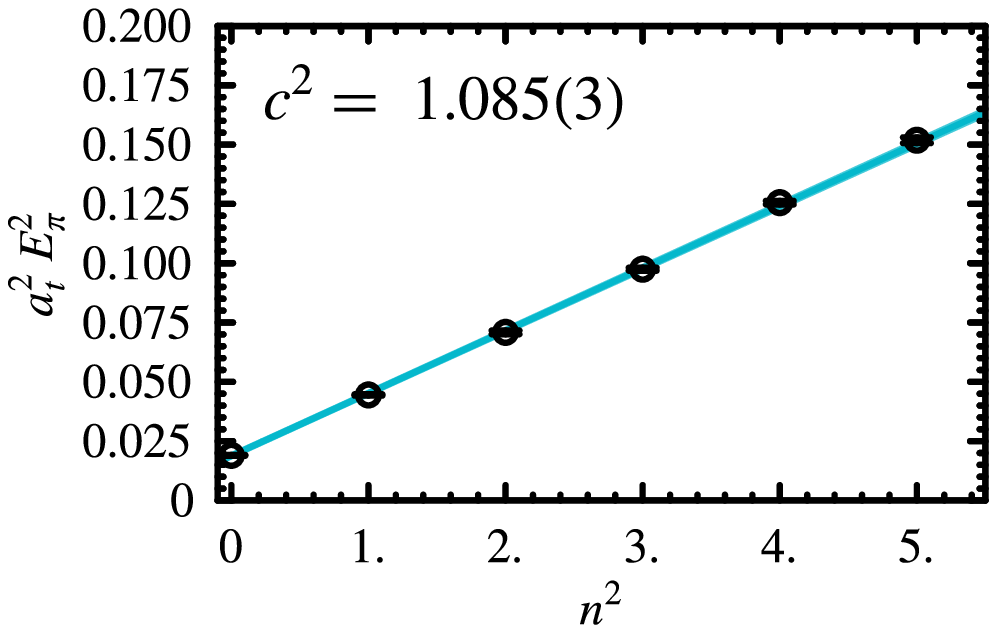}
\caption{Meson dispersion relation}\label{fig:dispersion}
\end{minipage}
\end{tabular}
\end{figure}

We search the remaining two-dimensional space by varying the parameters $\nu_s$ and $m_0$; the corresponding PCAC mass (in units of $a_t^{-1}$) is shown in Figure~\ref{fig:mst}. It is interesting to note that when $\nu_s$ is around 1 (the classic value), $M_s \approx M_t$ among range of different $m_0$ values. Figure~\ref{fig:m0-07} shows one of our parameters when $m_0$ is fixed close to the chiral limit (in our study, it is at $-0.07$, slightly below the critical mass). As we can see, at $\nu_s=1$ the $a_tM_s$ is automatically equal to $a_tM_t$. This suggests that with our setup (stout-smearing, tree-level tadpole coefficients for the fermion action) the tunings are consistent with the classic prediction.

One question is: how does our condition for the fermion anisotropy compare with the conventional dispersion relation ($c^2=(E^2-m^2)/p^2$) in a large volume? To demonstrate that the two are consistent at $O(a)$, we pick one of our simulation points, $m_0 = -0.054673$ and $\nu_s = 1.0$ (where $a_s = 0.116(3)$~fm), and we perform a dynamical simulation on a bigger volume, $12^3\times 128$ without Schr\"odinger functional scheme. We measure the pion dispersion relation (shown in Figure~\ref{fig:dispersion}), and we find $c^2=1.088(8)$. This is about the same amount of discrepancy as in the Schr\"odinger functional measurement of $M_s$ and $M_t$. Therefore, since it uses smaller volumes, probing the condition $M_s=M_t$ is a more efficient way to tune the fermion anisotropy $\nu_s$ than the dispersion relation and with Schr\"odinger functional, one can work on smaller pion masses (even near chiral limit) without too much additional cost.

The final check is: how good is our initial tadpole-improved $c_{\rm SW}^{\rm s,t}$? In the Schr\"odinger functional scheme, such nonperturbative coefficients are determined by requiring that
\vspace{-0.2cm}
\begin{eqnarray}
\Delta M = M(2T/4,T/4)  -  M^\prime(2T/4,T/4) = \Delta M^{{\rm Tree},M=0}
\end{eqnarray}
be satisfied. The tree-level $\Delta M$ value is obtained from simulation in a classical background field. In the dynamical simulation with parameter $\nu_s=1$ and $m_0=-0.056266$, we find $M_s=M_t$ is satisfied. (See Figure.~\ref{fig:mst} in which the thickness represents the number of the measurements done with that specific choice of parameters.) We further check the NP condition: the tree-level values of $a_s\Delta M$ are $-0.00056166$ and $-0.00028645$, and the measured $a_s\Delta M_{s,t}$ are $-0.000257(424)$ and $0.00009(12)$. Another case at parameters $\nu_s=0.9$ and $m_0=-0.069$ gives $M_s$ consistent with $M_t$ within the errorbar. The NP $c_{\rm SW}^{\rm s,t}$ condition is also satisfied. We found that tadpole-corrected tree-level coefficients with the stout-link smearing are consistent with the nonperturbative $O(a)$-improved coefficients in the three flavor dynamical simulation.

By imposing the condition of equality on the two PCAC masses ($M_t$, $M_s$) measured under the imposed background fields in two directions: $t$ and ``$z$'', we will find the correct parameters $\nu_s$ and $m_0$. A linear interpolation $M_{s,t}(\nu_s,m_0) = b_{s,t} + d_{s,t}\nu_s + e_{s,t}m_0$ would work for small variations in the parameter space. Some additional runs in the future will help us better determine these coefficients; a good interpolating form will help us when tuning the strange quark in the future.

\vspace{-0.2cm}
\section{Conclusion and Outlook}
\vspace{-0.4cm}
We demonstrate that the Schr\"odinger functional combined with stout-link smearing shows promise in dynamical runs. Further, we show in Schr\"odinger functional scheme that stout-link smearing and nonperturbatively modified tadpole factors automatically make our $O(a)$-improved coefficient $c_{\rm SW}^{s,t}$ tuning condition fulfilled. Our proposed finite-box fermionic anisotropy $\nu_s$ tuning (using the ratio of the PCAC mass measured with background field in space and time directions) is as good as conventional large-box runs but more efficient.

Our generation of two-flavor anisotropic ($\xi_R = 3$) Wilson fermion configurations is complete. This includes two lattice sizes: $L \approx 1.8, 2.6$~fm with $m_\pi \approx 400, 570$~MeV. In the near future, we will begin to fine tune the strange quark mass. We will also calculate $O(a)$-improved coefficients: $c_{V,A}$ and $Z_{V,A}$ for people who are interested in using these configurations for other physics.

\vspace{-0.4cm}
\section*{Acknowledgements}
\vspace{-0.4cm}
This work was done using the Chroma software suite\cite{Edwards:2004sx} on clusters at Jefferson Laboratory using time awarded under the SciDAC Initiative. Authored by Jefferson Science Associates, LLC under U.S. DOE Contract No. DE-AC05-06OR23177. The U.S. Government retains a non-exclusive, paid-up, irrevocable, world-wide license to publish or reproduce this manuscript for U.S. Government purposes.


\vspace{-0.4cm}
\begin{thebibliography}{10}
\expandafter\ifx\csname bibnamefont\endcsname\relax
  \def\bibnamefont#1{#1}\fi
\expandafter\ifx\csname bibfnamefont\endcsname\relax
  \def\bibfnamefont#1{#1}\fi
\expandafter\ifx\csname url\endcsname\relax
  \def\url#1{\texttt{#1}}\fi
\expandafter\ifx\csname urlprefix\endcsname\relax\def\urlprefix{URL }\fi
\expandafter\ifx\csname bibinfo\endcsname\relax \def\bibinfo#1#2{#2}\fi
\expandafter\ifx\csname eprint\endcsname\relax \def\eprint#1{#1}\fi
\vspace{-0.2cm}

\bibitem{all-roper}
\bibinfo{author}{\bibfnamefont{B.~G.}~\bibnamefont{Lasscock}} \emph{et~al.},
\bibinfo{journal}{arXiv:0705.0861 [hep-lat]},
\bibinfo{pages}{and references within}.

\bibitem{Mathur:2003zf}
\bibinfo{author}{\bibfnamefont{N.}~\bibnamefont{Mathur}} \emph{et~al.},
  \bibinfo{journal}{Phys. Lett.} \textbf{\bibinfo{volume}{B605}},
  \bibinfo{pages}{137} (\bibinfo{year}{2005}), \eprint{hep-ph/0306199}.

\bibitem{Morningstar:1999rf}
\bibinfo{author}{\bibfnamefont{C.~J.} \bibnamefont{Morningstar}}
  \bibnamefont{and} \bibinfo{author}{\bibfnamefont{M.~J.}
  \bibnamefont{Peardon}}, \bibinfo{journal}{Phys. Rev.}
  \textbf{\bibinfo{volume}{D60}}, \bibinfo{pages}{034509}
  (\bibinfo{year}{1999}), \eprint{hep-lat/9901004}.

\bibitem{Basak:2006ww}
\bibinfo{author}{\bibfnamefont{S.}~\bibnamefont{Basak}} \emph{et~al.}
  (\bibinfo{year}{2006}), \eprint{hep-lat/0609052}.

\bibitem{Harada:2001ei}
\bibinfo{author}{\bibfnamefont{J.}~\bibnamefont{Harada}},
  \bibinfo{author}{\bibfnamefont{A.~S.} \bibnamefont{Kronfeld}},
  \bibinfo{author}{\bibfnamefont{H.}~\bibnamefont{Matsufuru}},
  \bibinfo{author}{\bibfnamefont{N.}~\bibnamefont{Nakajima}}, \bibnamefont{and}
  \bibinfo{author}{\bibfnamefont{T.}~\bibnamefont{Onogi}},
  \bibinfo{journal}{Phys. Rev.} \textbf{\bibinfo{volume}{D64}},
  \bibinfo{pages}{074501} (\bibinfo{year}{2001}).

\bibitem{Aoki:2001ra}
\bibinfo{author}{\bibfnamefont{S.}~\bibnamefont{Aoki}},
  \bibinfo{author}{\bibfnamefont{Y.}~\bibnamefont{Kuramashi}},
  \bibnamefont{and} \bibinfo{author}{\bibfnamefont{S.-i.}
  \bibnamefont{Tominaga}}, \bibinfo{journal}{Prog. Theor. Phys.}
  \textbf{\bibinfo{volume}{109}}, \bibinfo{pages}{383} (\bibinfo{year}{2003}),
  \eprint{hep-lat/0107009}.

\bibitem{Hashimoto:2003fs}
\bibinfo{author}{\bibfnamefont{S.}~\bibnamefont{Hashimoto}} \bibnamefont{and}
  \bibinfo{author}{\bibfnamefont{M.}~\bibnamefont{Okamoto}},
  \bibinfo{journal}{Phys. Rev.} \textbf{\bibinfo{volume}{D67}},
  \bibinfo{pages}{114503} (\bibinfo{year}{2003}), \eprint{hep-lat/0302012}.

\bibitem{Umeda:2003pj}
\bibinfo{author}{\bibfnamefont{T.}~\bibnamefont{Umeda}} \emph{et~al.}
  (\bibinfo{collaboration}{CP-PACS}), \bibinfo{journal}{Phys. Rev.}
  \textbf{\bibinfo{volume}{D68}}, \bibinfo{pages}{034503}
  (\bibinfo{year}{2003}), \eprint{hep-lat/0302024}.

\bibitem{Morrin:2006tf}
\bibinfo{author}{\bibfnamefont{R.}~\bibnamefont{Morrin}},
  \bibinfo{author}{\bibfnamefont{A.~O.} \bibnamefont{Cais}},
  \bibinfo{author}{\bibfnamefont{M.}~\bibnamefont{Peardon}},
  \bibinfo{author}{\bibfnamefont{S.~M.} \bibnamefont{Ryan}}, \bibnamefont{and}
  \bibinfo{author}{\bibfnamefont{J.-I.} \bibnamefont{Skullerud}},
  \bibinfo{journal}{Phys. Rev.} \textbf{\bibinfo{volume}{D74}},
  \bibinfo{pages}{014505} (\bibinfo{year}{2006}).

\bibitem{Morningstar:2003gk}
\bibinfo{author}{\bibfnamefont{C.}~\bibnamefont{Morningstar}} \bibnamefont{and}
  \bibinfo{author}{\bibfnamefont{M.~J.} \bibnamefont{Peardon}},
  \bibinfo{journal}{Phys. Rev.} \textbf{\bibinfo{volume}{D69}},
  \bibinfo{pages}{054501} (\bibinfo{year}{2004}), \eprint{hep-lat/0311018}.

\bibitem{Edwards:2007}
\bibinfo{author}{\bibfnamefont{R.~G.} \bibnamefont{Edwards}},
  \bibinfo{author}{\bibfnamefont{B.}~\bibnamefont{Jo\'o}}, \bibnamefont{and}
  \bibinfo{author}{\bibfnamefont{H.-W.} \bibnamefont{Lin}},
  \bibinfo{journal}{In preparation} .

\bibitem{Chen:2000ej}
\bibinfo{author}{\bibfnamefont{P.}~\bibnamefont{Chen}}, \bibinfo{journal}{Phys.
  Rev.} \textbf{\bibinfo{volume}{D64}}, \bibinfo{pages}{034509}
  (\bibinfo{year}{2001}), \eprint{hep-lat/0006019}.

\bibitem{alpha}
\bibinfo{author}{\bibfnamefont{M.}~\bibnamefont{Luscher}},
  \bibinfo{author}{\bibfnamefont{R.}~\bibnamefont{Narayanan}},
  \bibinfo{author}{\bibfnamefont{P.}~\bibnamefont{Weisz}}, \bibnamefont{and}
  \bibinfo{author}{\bibfnamefont{U.}~\bibnamefont{Wolff}},
  \bibinfo{journal}{Nucl. Phys.} \textbf{\bibinfo{volume}{B384}},
  \bibinfo{pages}{168} (\bibinfo{year}{1992});
\bibinfo{author}{\bibfnamefont{M.}~\bibnamefont{Luscher}},
  \bibinfo{author}{\bibfnamefont{S.}~\bibnamefont{Sint}},
  \bibinfo{author}{\bibfnamefont{R.}~\bibnamefont{Sommer}}, \bibnamefont{and}
  \bibinfo{author}{\bibfnamefont{P.}~\bibnamefont{Weisz}},
  \bibinfo{journal}{Nucl. Phys.} \textbf{\bibinfo{volume}{B478}},
  \bibinfo{pages}{365} (\bibinfo{year}{1996}); 
\bibinfo{author}{\bibfnamefont{M.}~\bibnamefont{Luscher}},
  \bibinfo{author}{\bibfnamefont{S.}~\bibnamefont{Sint}},
  \bibinfo{author}{\bibfnamefont{R.}~\bibnamefont{Sommer}},
  \bibinfo{author}{\bibfnamefont{P.}~\bibnamefont{Weisz}}, \bibnamefont{and}
  \bibinfo{author}{\bibfnamefont{U.}~\bibnamefont{Wolff}},
  \bibinfo{journal}{Nucl. Phys.} \textbf{\bibinfo{volume}{B491}},
  \bibinfo{pages}{323} (\bibinfo{year}{1997}).

\bibitem{Jansen:1998mx}
\bibinfo{author}{\bibfnamefont{K.}~\bibnamefont{Jansen}} \bibnamefont{and}
  \bibinfo{author}{\bibfnamefont{R.}~\bibnamefont{Sommer}}
  (\bibinfo{collaboration}{ALPHA}), \bibinfo{journal}{Nucl. Phys.}
  \textbf{\bibinfo{volume}{B530}}, \bibinfo{pages}{185} (\bibinfo{year}{1998}),
  \eprint{hep-lat/9803017}.

\bibitem{Yamada:2004ja}
\bibinfo{author}{\bibfnamefont{N.}~\bibnamefont{Yamada}} \emph{et~al.}
  (\bibinfo{collaboration}{JLQCD}), \bibinfo{journal}{Phys. Rev.}
  \textbf{\bibinfo{volume}{D71}}, \bibinfo{pages}{054505}
  (\bibinfo{year}{2005}), \eprint{hep-lat/0406028}.

\bibitem{Aoki:2005et}
\bibinfo{author}{\bibfnamefont{S.}~\bibnamefont{Aoki}} \emph{et~al.}
  (\bibinfo{collaboration}{CP-PACS}), \bibinfo{journal}{Phys. Rev.}
  \textbf{\bibinfo{volume}{D73}}, \bibinfo{pages}{034501}
  (\bibinfo{year}{2006}), \eprint{hep-lat/0508031}.

\bibitem{Edwards:1997nh}
\bibinfo{author}{\bibfnamefont{R.~G.} \bibnamefont{Edwards}},
  \bibinfo{author}{\bibfnamefont{U.~M.} \bibnamefont{Heller}},
  \bibnamefont{and} \bibinfo{author}{\bibfnamefont{T.~R.}
  \bibnamefont{Klassen}}, \bibinfo{journal}{Phys. Rev. Lett.}
  \textbf{\bibinfo{volume}{80}}, \bibinfo{pages}{3448} (\bibinfo{year}{1998}),
  \eprint{hep-lat/9711052}.

\bibitem{Klassen:1998ua}
\bibinfo{author}{\bibfnamefont{T.~R.} \bibnamefont{Klassen}},
  \bibinfo{journal}{Nucl. Phys.} \textbf{\bibinfo{volume}{B533}},
  \bibinfo{pages}{557} (\bibinfo{year}{1998}), \eprint{hep-lat/9803010}.

\bibitem{Edwards:2004sx}
\bibinfo{author}{\bibfnamefont{R.~G.} \bibnamefont{Edwards}} \bibnamefont{and}
  \bibinfo{author}{\bibfnamefont{B.}~\bibnamefont{Joo}}
  (\bibinfo{collaboration}{SciDAC}), \bibinfo{journal}{Nucl. Phys. Proc.
  Suppl.} \textbf{\bibinfo{volume}{140}}, \bibinfo{pages}{832}
  (\bibinfo{year}{2005}), \eprint{hep-lat/0409003}.

\end{thebibliography}
\end{document}